\documentclass[12pt]{article}
\usepackage{graphicx}
\usepackage{dcolumn}
\usepackage{amssymb}
\usepackage{bm}
\usepackage{enumerate}
\begin{document}
\title{\bf A strong astrophysical constraint on the
  violation of special relativity by quantum
  gravity\footnote{Previous title:
 {\em Lorentz violation and Crab synchrotron emission:
a new constraint far beyond the Planck scale}}}
\author{T.~Jacobson,
S.~Liberati,
D.~Mattingly
\\[2mm]
{\small \it
Department of Physics,}\\
{\small \it University of Maryland, College Park, MD 20742-4111,
USA.} }
\date{}
\maketitle


\vfill
\def\wt{\widetilde}
\def\gsim{\; \raisebox{-.8ex}{$\stackrel{\textstyle >}{\sim}$}\;}
\def\lsim{\; \raisebox{-.8ex}{$\stackrel{\textstyle <}{\sim}$}\;}
\def\half{{1\over2}}
\def\a{\alpha}
\def\b{\beta}
\def\g{\gamma}
\def\e{\epsilon}
\def\m{\mu}
\def\o{\omega}
\def\L{{\mathcal L}}
\def\d{{\mathrm{d}}}
\def\p{{\mathbf{p}}}
\def\q{{\mathbf{q}}}
\def\k{{\mathbf{k}}}
\def\fp{{p_{\rm 4}}}
\def\fq{{q_{\rm 4}}}
\def\fk{{k_{\rm 4}}}
\def\etal{{\emph{et al}}}
\def\det{{\mathrm{det}}}
\def\tr{{\mathrm{tr}}}
\def\ie{{\emph{i.e.}}}
\def\aka{{\emph{aka}}}

{\bf Special relativity asserts that physical phenomena
  appear the same for all inertially moving
  observers. This symmetry, called Lorentz symmetry,
  relates long wavelengths to short ones: if the symmetry
  is exact it implies that spacetime must look the same
  at all length scales. Several approaches to quantum
  gravity, however, suggest that there may be a Lorentz
  violating microscopic structure of spacetime, for
  example discreteness~\cite{loopqg},
  non-commutativity~\cite{Carroll:2001ws}, or extra
  dimensions~\cite{Burgess:2002tb}. Here we determine a
  very strong constraint on a type of Lorentz violation
  that produces a maximum electron speed less than the
  speed of light. We use the observation of 100 MeV
  synchrotron radiation from the Crab nebula to improve
  the previous limits by a factor of 40 million, ruling
  out this type of Lorentz violation, and thereby
  providing an important constraint on theories of
  quantum gravity.}

To characterize Lorentz violation we adopt the simple framework
of deformed dispersion relations. A dispersion relation is the
relation between energy $E$ and momentum $p$ for a particle. In
relativity, these quantities change under a Lorentz
transformation, but the particle dispersion relation $E^2 = m^2c^2
+ p^2c^4$ is invariant (where $m$ is the particle mass
and $c$ is the speed of light). Lorentz transformations include both
``boosts", which are transformations to a relatively moving
frame, and rotations.

We consider dispersion relations  that break boost invariance but
preserve rotation invariance. (A dispersion relation that is not
boost invariant can hold in only one frame. We assume this frame
coincides with that of the cosmic microwave background.)  In
particular for photons and electrons we consider the dispersion
relations
\begin{eqnarray}
\omega^2(k)&=& k^2+\xi \frac{k^3}{M},
\label{eq:pdr}\\
E^2(p)&=& m^2+ p^2+\eta \frac{p^3}{M}, \label{eq:mdr}
\end{eqnarray}
where $\omega$ and $k$ are the photon frequency and wave
number, and $E$ and $p$ are the electron energy and
momentum. We use units with Planck's constant $\hbar$ and
the low energy speed of light $c$ both set to unity (thus
the photon energy $\hbar\o$ is also denoted by $\o$). The
energy scale $M=10^{19}$ GeV, which is close to the
Planck energy, $M_{\rm Planck}=1.22\times 10^{19}$ GeV, is
factored out explicitly. We regard these relations as
phenomenological, not fundamental. In particular one
would expect a whole series of higher order terms in the
momentum, but those would be negligible at the energies
we consider. Extra lower order terms are already strongly
enough constrained by observation so as to be irrelevant
for the new constraint considered here.

Dispersion relations with higher order momentum
corrections analogous to (\ref{eq:pdr}) are familiar from
common physical situations like vibrational waves in a
crystal or light waves in a refractive medium, where the
long wavelength (low momentum) modes travel at a common
speed, while the modes whose wavelength is short enough
to be sensitive to the microscopic structure of the
medium behave differentially. Several approaches to
quantum gravity propose a Lorentz violating microscopic
structure of spacetime.  There is no unique prediction
regarding the values of the dimensionless coefficients
$\xi$ and $\eta$, but if not suppressed by some other
symmetry they would presumably be of order unity if they
originate from quantum gravity effects. This is because
physics rarely produces dimensionless numbers that differ
by many orders of magnitude from unity from theories with
only one scale, and the unique energy scale $M$ of
quantum gravity has already been factored out in
(\ref{eq:pdr},\ref{eq:mdr}).

It is of course possible that other symmetries along with
boost symmetry are violated.   However, a bound on pure boost
violation is also a bound on such multiple symmetry violations,
barring an unlikely cancellation of different effects.  Hence, to
keep the analysis simple we assume all of standard physics except
boost invariance. In particular we preserve rotation symmetry and
electromagnetic gauge invariance, additivity of energy and
momentum for multi-particle systems, and energy-momentum
conservation in particle interactions.

A consistent dynamical framework that yields modified dispersion
and preserves our other assumptions is effective field theory.
Effective field theory is very general and can incorporate Lorentz
violation arising from a wide range of underlying quantum gravity
scenarios, including for example string theory and spacetime
discreteness. It has been shown~\cite{MP} in effective field
theory that left and right polarizations of photons have opposite
values of $\xi$ in (\ref{eq:pdr}), while left and right electron
chiralities can have independent $\eta$ values. Polarization
dependence of $\xi$ is unimportant for the new constraint
described here, although we shall use it at the end when
combining our new constraint with previous work. We assume here
that $\eta$ is not chirality dependent.

Perhaps surprisingly, Planck energies are not needed to get
constraints on $\xi$ and $\eta$ of order unity or even less. For
example, the group velocities $d\o/dk$ and $dE/dp$ for photons
and electrons satisfying equations (\ref{eq:pdr},\ref{eq:mdr})
depend on these parameters and on the particle energy.
%
%
Over long propagation distances this could produce observable
arrival time differences for simultaneously emitted photons of
different energies~\cite{Pav,Amelino-Camelia:1997gz}. Arrival
time constraints have been obtained using the observed radiation
from gamma ray bursts~\cite{schaefer}, active galactic
nuclei~\cite{Biller}, and pulsars~\cite{Kaaret:1999ve}. The best
that can be done with the current data is $|\xi|\lesssim {\rm
O}(100)$~\cite{schaefer,Biller} but more stringent limits can be
expected in the future~\cite{Norris:1999nh}. Constraints of order
$10^{-4}$ have been obtained on the {\it difference} in $\xi$ for
different polarizations (birefringence) using spectropolarimetric
observations of distant galaxies~\cite{GK}. Using the effective
field theory result of \cite{MP} this yields the powerful new
constraint $|\xi|\lesssim O(10^{-4})$.

Complementary constraints were recently achieved by considering
the threshold reactions of photon decay, vacuum \v{C}erenkov
radiation, and photon absorption involving high energy photons
and electrons~\cite{CG,Gonzalez-Mestres:1995jg,Stecker:2001vb,
Jacobson:2001tu,JLM02,Konopka:2002tt,Amelino-Camelia:2002dx,Jacobson:2003ty}.
Energies around 10 TeV are needed in order to put constraints of
order unity on the deformation parameters for electrons and
photons using threshold effects. Energies higher than this are
achieved in astrophysical sources and hence threshold reactions
place at least order unity bounds on $\eta,\xi$ .

However, If we are to rule out quantum gravity effects
characterized by (\ref{eq:pdr},\ref{eq:mdr}) it
is necessary to find observations that limit $\eta$ and
$\xi$ to be much smaller than unity. None of the previous
work has accomplished this for both electrons and
photons. Here we show that the presence of
synchrotron radiation from the Crab nebula, emitted by
electrons cycling in a magnetic field, provides the new
constraint $\eta>-7\times 10^{-8}$ on the electron
parameter $\eta$ within the effective field theory
framework.  This improves previous constraints by a
very large amount. In particular it implies that the
parameter $E_{QG}$ of the popular phenomenological model
of Ref.~\cite{Amelino-Camelia:1997gz} (which corresponds
to $M/|\eta|$, with $\xi=\eta$ in our notation) is
bounded by $E_{QG}> 10^7\, M_{\rm Planck}$. (To be
compatible with effective field theory that model must be
modified to accommodate the polarization-dependence of
the sign of $\xi$.) This is seven orders of magnitude
stronger than the constraint from photon absorption
$E_{QG}> 0.3\, M_{\rm
  Planck}$~\cite{Jacobson:2001tu,JLM02,Stecker:2003ni}
(see also \cite{Amelino-Camelia:2002dx,Jacobson:2003ty}
for a debate on the validity of this constraint) and
three orders of magnitude stronger than the new
birefringence constraint of \cite{MP}.


Synchrotron radiation in the Crab nebula is produced by electrons
cycling in a magnetic field of around 0.6 mG. To produce the
observed radiation of energy 100 MeV in this field requires in
standard electrodynamics a gamma factor $\g=(1-v^2)^{-1/2}$ of
$3\times 10^{9}$, corresponding to an electron energy of 1500
TeV. To achieve this energy the electron velocity must differ from
the speed of light by less than one part in $10^{19}$.  Lorentz
violation with negative $\eta$ puts an upper bound on the electron
speed and hence an upper bound on the possible synchrotron
radiation frequency. (The existence of a cutoff
  in the synchrotron radiation at high charged particle
  energies in the presence of Lorentz violation was
  previously noted in Ref.~\cite{Gonzalez-Mestres:2000eh}.)
But to apply this reasoning we must first re-analyze the
synchrotron emission process allowing for the Lorentz violating
effects that we wish to bound.

In standard electrodynamics, accelerated electrons in a magnetic
field emit synchrotron radiation with a spectrum that sharply cuts
off at a frequency $\omega_c$ given by the formula~\cite{Jackson}
\begin{equation}
 \omega_c^{\rm LI}=\frac{3}{2}\frac{eB\gamma^2}{m},
  \label{eq:opeak}
\end{equation}
where $B$ is the component of the magnetic field
orthogonal to the electron path. The gamma factor
$\gamma=E/m$ grows without bound with the electron energy
$E$ so there is no upper limit on $\omega_c^{\rm LI}$.

Equation (\ref{eq:opeak}) is based on the electron trajectory
in a given magnetic field, the radiation produced by a given
current, and the relativistic relation between energy and
velocity, all of which could be affected by Lorentz violation. We
find that in the presence of Lorentz violation, (\ref{eq:opeak})
becomes
\begin{equation}
\omega_c=\frac{3}{2} \frac{eB}{m}\frac{m\gamma(E)}{E}\;
\gamma^2(E), \label{eq:opeaklv}
\end{equation}
where $\gamma(E)=\gamma(v(E))$. (For derivations of
(\ref{eq:opeaklv}) and the constraint (\ref{eq:synchcon})
below see the Methods section.) If $\eta$ is negative,
electrons have a maximal velocity less than $c$, hence
$\gamma(E)$ is a bounded function of $E$.  This implies
that there is a maximum achievable value of the cutoff
frequency as given by equation (\ref{eq:opeaklv}), which
we denote by $\omega_c^{\rm max}$. The value of the
energy that yields this cutoff frequency is higher than
the Lorentz-invariant value.

The rapid decrease in amplitude of synchrotron emission
at frequencies larger than $\omega_c$ implies that most
of the flux at a given frequency in a synchrotron
spectrum is due to electrons whose $\omega_c$ is above
that frequency. Thus $\omega_c^{\rm max}$ must be greater
than the maximum observed synchrotron emission frequency
$\o_{\rm obs}$. This yields the constraint
\begin{equation}
\eta > - \frac{M}{m}\left(\frac{0.34\, eB}{m\o_{\rm
obs}}\right)^{3/2}. \label{eq:synchcon}
\end{equation}
%


The strongest synchrotron constraint comes from the
system that minimizes the ratio $B/\o_{\rm obs}$, which
turns out to be the Crab nebula. The Crab nebula, a
supernova remnant (SNR), is a bright source of radio,
optical, X-ray and gamma-ray emission, which exhibits a
broad spectrum characterized by two marked humps. This
spectrum is consistently explained by a combination of
electron synchrotron emission and inverse Compton
scattering of ambient photons by high energy
electrons~\cite{AA96, deJager}. In fact, no other model
for the emission is under consideration, other than for
producing some of the highest energy photons. For our
constraint we assume that this standard SNR model is
correct.

The Crab synchrotron emission has been observed to extend
at least up to energies of about 100 MeV~\cite{AA96,
  deJager}, just before the inverse Compton hump begins
to contribute to the spectrum. The magnetic field in the
emission region has been estimated by several methods
which agree on a value between 0.15--0.6 mG (see
e.g.~\cite{Hillas} and references therein.) Two of these
methods, radio synchrotron emission and equipartition of
energy, are insensitive to Planck suppressed Lorentz
violation, hence we are justified in adopting a value of
this order for the purpose of constraining Lorentz
violation. We shall use the largest value 0.6 mG for $B$
since it yields the weakest constraint (Eq.
(\ref{eq:synchcon}) shows how the constraint scales with
$B$).

Using the above values for the magnetic field and the highest
synchrotron  frequency, Eq.~(\ref{eq:synchcon}) yields the lower
bound
\begin{equation}
\eta > - 7\times 10^{-8}.\label{eq:numba2}
\end{equation}
This corresponds to
\begin{equation}
E_{\rm QG}=\frac{M}{|\eta|} > 10^{26}\, {\rm GeV}
\label{eq:numbaQG}
\end{equation}
in the phenomenological framework of
Ref.~\cite{Amelino-Camelia:1997gz}. Thus $E_{\rm QG} $ is
constrained to be at least seven orders of magnitude
larger than the Planck energy.

To complement the synchrotron constraint, which just
bounds $\eta$ from below, one can use the vacuum
birefringence and \v{C}erenkov constraints. Lack of
observed vacuum birefringence bounds the difference in
$\xi$ for right and left circular polarized
photons~\cite{GK}. Using the effective field theory
result of \cite{MP} that left and right circular
polarized photons have opposite values for $\xi$, this
yields the constraint $|\xi|< 4\times10^{-4}$.

To bound $\eta$ from above we can use the vacuum
\v{C}erenkov constraint. As noted by previous
authors~\cite{Gonzalez-Mestres:1995jg,CG,Jacobson:2001tu,Konopka:2002tt},
modified dispersion can cause electrons to emit
\v{C}erenkov radiation in vacuum and rapidly lose
energy. It was suggested in~\cite{Stecker:2003ni} that a
very strong \v{C}erenkov constraint is obtained for
positive $\eta$ using the high energy electrons that
produce the synchrotron radiation. But for positive
$\eta$ the energy required to produce a given cutoff
synchrotron frequency goes {\it down}, as the electron
group velocity can increase beyond the speed of light and
$\gamma(E)$ can diverge at finite energy.  Hence the
electron energy may be much lower than it is in the
Lorentz-invariant case. (Nevertheless a constraint can be
obtained in this way, as we shall show in a future
publication with F.~Stecker.) Thus we look to other high
energy electrons whose existence is not in question.

The inverse Compton peak in the Crab spectrum contains
energies up to 50 TeV.  By energy conservation, this
implies that electrons of at least 50 TeV propagate,
hence the values of $\eta$ and $\xi$ must not allow these
electrons to emit vacuum \v{C}erenkov radiation. Within
the region not ruled out by the birefringence bound, the
results of~\cite{Jacobson:2001tu,JLM02,Konopka:2002tt}
yield $\eta<0.01$.

Finally, recall that we neglected $\xi$ in obtaining the
constraint. Using Eq. (\ref{eq:xineglect}) one can see that this
is justified everywhere in the region not already excluded by the
absence of vacuum \v{C}erenkov radiation. The worst case would be
at the smallest $|\eta|$. With $\eta$ given by the lower bound
(\ref{eq:numba2}), Eq.~(\ref{eq:xineglect}) shows that $\xi$ can
be neglected provided $|\xi|\lesssim 30$, which is a factor of
nearly $10^5$ larger than the vacuum birefringence limit.

Putting these observations together, we conclude that Lorentz
violation suppressed by the ratio $E/E_{\rm Planck}$ and
compatible with effective field theory is constrained in all
directions of the $\xi$-$\eta$ parameter space by much stronger
than order unity bounds. The allowed region is a skinny rectangle,
with bounds from vacuum birefringence above and below by $|\xi|<
4\times 10^{-4}$, from Crab synchrotron radiation on the left by
$\eta>- 7\times 10^{-8}$, and from vacuum \v{C}erenkov radiation
on the right by $\eta<0.01$.

It is notable that, in such a short time after the
recognition of the possibility of obtaining interesting
bounds on Planck suppressed Lorentz violation, it has
been so severely constrained. The constraint reported
here on a possible maximal electron speed less than the
speed of light is more than seven orders of magnitude
better than previous limits.  This limit is strong enough
to conclude that quantum gravity scenarios postulating or
implying this sort of Lorentz violation are not viable.

The type of Lorentz violation we have considered also
violates CPT symmetry\cite{MP}, the symmetry under
combined charge conjugation, space inversion, and time
reversal. The fact that it is ruled out might therefore
be a consequence of CPT symmetry, rather than Lorentz
symmetry. It thus makes sense to think seriously about
constraining Lorentz violation suppressed by the second
power, $(E/E_{\rm Planck})^2$. Some interesting bounds
already exist at this order~\cite{Aloisio:2000cm,JLM02},
and other possibilities have been
discussed\cite{JLM02,Amelino-Camelia:2003bt}. If it were
eventually possible to achieve strong bounds at this
order then it would be reasonable to conclude that
Lorentz symmetry is simply not violated, since there is
no other symmetry that could protect against this sort of
Lorentz violation. Alternatively, if the search for such
constraints finds a Lorentz violation, that would open up
a much needed observational window on the difficult
problem of quantum gravity.

\section*{Methods}

To determine the effects of Lorentz violation of
(\ref{eq:opeak}) we follow the heuristic derivation of
(\ref{eq:opeak}) given in Ref.~\cite{Jackson} assuming
the framework of effective field theory. Without assuming
Lorentz invariance, the {\it purely kinematical} result
is
\begin{equation}
\omega_c=\frac{3}{4} \frac{1}{R(E)\delta(E)}\;
\frac{1}{c(\omega_c)-v(E)} \label{eq:opeaktilde}
\end{equation}
where $R(E)$ is the radius of curvature of the orbit,
$\delta(E)$ is the opening angle for the forward-directed
radiation pattern, and $c(\omega_c)$ and $v(E)$ are the
group velocities of the radiation and electron
respectively. The solution of Eq.~(\ref{eq:opeaktilde})
for $\omega_c(E)$ determines the cutoff synchrotron
frequency. In Eq. (\ref{eq:opeaktilde}) we have used the
fact that the electron and photon speeds are very close
to the low energy speed of light $c$, which is set equal
to unity. The numerical constant is chosen to yield the
correct relativistic result (\ref{eq:opeak}) with
$\delta(E)=\gamma^{-1}(E)$ (where $\g(E)=\g(v(E))$).

The radius $R(E)$ for a given energy is determined by the
equation of motion of the electron in a magnetic
field. All Lorentz violating terms in the equation of
motion are suppressed by $M^{-1}$, and the leading high
energy corrections come from modifications to the minimal
coupling terms. To estimate the magnitude of the change
we use the dispersion relation as the Hamiltonian. As
usual the minimal coupling is incorporated replacing the
momentum by ${\bf p} - e{\bf A}$, where ${\bf A}$ is a
vector potential for the magnetic field. This yields the
equation of motion ${\bf a}=[1 + 3\eta E/2M](e/E)\, {\bf
  v}\times{\bf B}$, where we have kept only the lowest
order term in $\eta$ and assumed relativistic energy
$E\gg m$. Since $E\ll M$, the presence of the Lorentz
violation makes very little difference to the orbital
equation, hence we conclude that to a very good
approximation the radius is related to the magnetic field
and the energy of the electron by the standard formula
$R(E)= E/eB$ (where again the speed of the electron has
been set equal to unity).

The angle $\delta(E)$ scales in the Lorentz invariant
case as $\gamma^{-1}(E)$. Since $R(E)$ and hence the
charge current is nearly unaffected by the Lorentz
violation, any significant deviation in $\delta(E)$ can
only come from the modified response of the
electromagnetic field to a given current.  The amplitude
of the modification is suppressed by at least one power
of $M$ however, so dimensional analysis implies that it
is suppressed by something of order $\xi \o/M$. For the
Crab synchrotron radiation used in our constraint $\o$ is
100 MeV, which yields a suppression factor of
$10^{-20}\xi$. Therefore the amplitude of the
electromagnetic field modification is negligible in
comparison to the zeroth order synchrotron radiation
(which is still present), and hence $\delta(E)$ scales
with $\gamma^{-1}(E)$ in the usual way.


Since the emitted photons have relatively low energy
compared to the electrons, it turns out that $\xi$ can be
neglected in the relevant region of parameter
space. Thus, since $v(E)$ is very close to $c$, the
reciprocal of the difference of group velocities in the
last term of Eq.~(\ref{eq:opeaktilde}) is well
approximated by $2\gamma^2(E)$. This yields
(\ref{eq:opeaklv}).

The maximum synchrotron frequency $\omega_c^{\rm max}$ is
obtained by maximizing $\o_c$ (\ref{eq:opeaklv}) with
respect to the electron energy, which amounts to
maximizing $\gamma^3(E)/E$.  The difference of group
velocities is given by
\begin{equation}
 c(\omega)-v(E)= \xi\, \frac{\omega}{M} + \frac{m^2}{2E^2} -
 \eta\, \frac{E}{M},
 \label{eq:vdiff}
\end{equation}
where we have used the dispersion relations
(\ref{eq:pdr},\ref{eq:mdr}) and dropped higher order terms.
Dropping the $\xi$ term this yields
\begin{equation}
 \gamma(E)\approx
  \left(\frac{m^2}{E^2}-2\eta\frac{E}{M} \right)^{-1/2},
   \label{eq:gm}
\end{equation}

Using this expression we find
\begin{equation}
\omega_c^{\rm max}=0.34 \, \frac{eB}{m}(-\eta
m/M)^{-2/3},\label{eq:opeaklv2}
\end{equation}
which is attained at the energy $E_{\rm
  max}=(-2m^2M/5\eta)^{1/3}=10\, (-\eta)^{-1/3}$
TeV. Since $\eta$ is negative this is higher than the
Lorentz invariant value that produces the same frequency,
but only by a factor of order unity that works out to be
$(9/5)^{3/4}\approx 1.55$. The energy of the electrons
that produce the synchrotron radiation of frequency 100
MeV is 1500 TeV in the Lorentz invariant case. If this
were the maximum frequency in the Lorentz violating case,
it would be produced by $\sim 2300$ TeV electrons.

We now analyze the question of neglecting $\xi$ in
evaluating the difference of group velocities
(\ref{eq:vdiff}). At the energy $E_{\rm max}$, the ratio
of the $\xi$-dependent term to the other terms is given
by
\begin{equation}
\frac{|\xi|\o}{(m^2M/2E^2) - \eta E} = 3\times 10^{-11}\,
|\xi|(-\eta)^{-4/3}.\label{eq:ratio}
\end{equation}
Hence neglecting $\xi$ is justified provided
\begin{equation}
|\xi|\lesssim 10^{11}(-\eta)^{4/3}. \label{eq:xineglect}
\end{equation}
%

\section*{Acknowledgements}

We wish to thank F.A.~Aharonian, G.E.~Allen, G. Amelino-Camelia,
and F. Stecker for helpful discussions. This work was supported in
part by the NSF.


\end{document}